\documentclass[aps, prx, reprint, longbibliography]{revtex4-2}
\usepackage{xr-hyper, hyperref, graphicx, amsmath, amssymb, xcolor,lineno}
\graphicspath{{figures/}}

\begin{document}
\title{Large Nernst effect in chemically derived multilayer graphene at millitesla magnetic fields}

\author{Valentin Semkin$^{1}$}
\email[]{semkin.va@phystech.edu}

\author{Denis Borisenko$^{1,2,3}$}

\author{Yana Litun$^{2}$}

\author{Oleg Kononenko$^{2,4}$}

\author{Dmitry Mylnikov$^{1}$}

\author{Alexey Bocharov$^{1,2}$}

\author{Dmitry Svintsov$^{1,2}$}
\email[]{svintcov.da@mipt.ru}

\affiliation{$^{1}$Moscow Institute of Physics and Technology, Dolgoprudny 141700, Russia}
\affiliation{$^{2}$Joint-Stock Company ''Skanda Rus'', Krasnogorsk 143403, Russia}
\affiliation{$^{3}$National Research Nuclear University “MEPhI”, Moscow 115409, Russia}
\affiliation{$^{4}$Institute of Microelectronics Technology and High Purity Materials, Russian Academy of Sciences, Chernogolovka 142432, Russia}

\begin{abstract}
Simple synthesis of graphitic compounds, their high conductivity, and integrability with other materials motivate the effort toward graphite-based thermoelectric generators. At the same time, semimetallic nature of graphite and graphene results in nearly-zero thermopower due to electron-hole compensation. Here, we observe large transverse thermopower in chemically derived multilayer graphene films with strong fluctuations of thickness and carrier density at low magnetic fields $B$. Using the scanning laser-induced heating of macroscopic film, we find that transverse (Nernst) thermoelectric voltage becomes comparable to the longitudinal thermoelectric voltage at the metal-doped graphene contact at $B^*\approx4$ mT and ambient conditions. Estimates of transverse thermopower $S_{xy}$ based on the known laser-induced temperature show that it is as large as $\sim 10$ $\mu$V/K at $B^*$, and raises in a sub-linear fashion to 250 $\mu$V/K at $B\approx315$ mT, the maximum field we reach with centimeter-sized permanent magnet. Extra increase in thermoelectric signal is achieved upon voltage measurement at Hall probes when the dc field lines are co-directional with local Nernst current. Our results show the promise of large-scale multilayer graphene for thermoelectricity generation.
\end{abstract}
\maketitle

Thermoelectricity enables generation of useful electric power from heat waste. The search for thermoelectric materials represents a compromise between large power factor ${\rm ZT}$ and fabrication ease~\cite{TE_materials_review}. Natural abundance of graphite, its electrical conductivity, ease of chemical synthesis and environmental stability have resulted in strong effort toward its thermoelectric applications~\cite{TE_Applications_graphene}. At the same time, graphite and multilayer graphene represent compensated electron-hole semimetals~\cite{Slonch_Weiss_bands,mcclure1958analysis}, which results in nearly-zero thermopower (Seebeck coefficient) $S$. The power factor of graphite and its thin-film variants, ${\rm ZT}=S^2\sigma T/\kappa$, where $\sigma$ and $\kappa$ are the electrical and thermal conductivities, is thus very low.

The methods to push graphitic films from charge neutrality and enhance ${\rm ZT}$ include metal-induced doping~\cite{TE_enhancement_metal_induced,Fe_doping_graphite,Gradient_doping_graphite}, intercalation  doping~\cite{Intercalation_doping_graphite}, nanomeshing and nanostructuring~\cite{TE_enhancement_nanomeshing}. It is impossible to guess the sign and magnitude of doping in these processes, and choosing the right chemistry is a matter of trial. Electrically induced doping via field effect represents a controllable strategy~\cite{Kim_thermopower}. However, the shift in Fermi level at given transverse electric field shrinks very quickly with increase in layer number~\cite{Negishi_Turbostraticity_on_mobility}. As a result, multilayer graphene with layer number above $\sim 10$ is hardly gateable. The thermopower of graphene can be increased by electron interactions~\cite{Interaction_thermopower,Zarenia_thermopower}, yet the effect occurs only in highest-quality structures.

It is possible to achieve large thermoelectric currents in compensated semimetals in a transverse magnetic field, a phenomenon known as the Nernst effect. The drift of electrons and holes in thermal gradient is co-directional, while the deflections of their velocity by magnetic field are opposite. Combined with opposite signs of charge, this leads to additive currents of electrons and holes under simultaneous action of thermal gradient and magnetic field. For a long time, the Nernst effect retained a merely academic phenomenon as it required either very large magnetic fields or low temperatures and high mobilities. Particularly, large Nernst effect was observed~\cite{Nernst_HOPG} in pyrolythic graphite films at temperatures $T<20$ K and mobility $\mu =3\times10^5$ cm$^2$/(V s), in exfoliated graphene at $T=40$ K and $B \sim 0.3$ T~\cite{Nernst_Graphene}, and in various Moire structures including oriented graphene on boron nitride at $T=1.7$ K~\cite{Elesin_Nernst} and twisted double graphene bilayer at $T\sim 1$ K and $B\sim 1$ T~\cite{Nernst_TDBG}. Stringent constraints on large Nernst coefficient promoted a search for anomalous Nernst effect without magnetic field~\cite{ANE_1,ANE_2,ANE_3}; the respective material platforms remain scarce.

While large Nernst coefficient ${\rm N}$ at charge neutrality is well-understood both for bulk semiconductors~\cite{InSb_Nernst,InSb_Nernst_1} and monolayer graphene~\cite{Kim_thermopower}, it remains unclear whether ${\rm N}$ reach large values in scalable chemical vapor deposited (CVD) multilayer graphene. Reports on synthesis of such films are numerous~\cite{Synt_Yang_Turbostratcic_Self_Heating,Synt_Garlow_TurbostraticMLG_Raman,Brzhezinskaya2021}, yet none of them measured the thermopower in magnetic field. Even the estimates of carrier mobility in CVD multilayers are scarce~\cite{Negishi_Turbostraticity_on_mobility} and yield values no larger than $\mu\lesssim2.5\times10^3$ cm$^2$/V s. Indeed, CVD synthesis can reduce mobility via grain boundaries and charged contaminants. The presence of a dimensionless prefactor $\mu B$ in the expression for the off-diagonal thermopower $S_{xy}$ deems large Nernst effect in such films unlikely. Another argument against strong Nernst effect comes from non-uniformity of thickness and local stacking in CVD multilayer graphene~\cite{Synt_Garlow_TurbostraticMLG_Raman}. At the same time, small apparent Hall mobility~\cite{Negishi_Turbostraticity_on_mobility} can arise from electron-hole compensation in multilayer graphene. The true mobility (e.g. measured via magnetoresistance) can be much larger, as reported by one of us for CVD multilayer graphene~\cite{Brzhezinskaya2021,Kononenko2022} and independently for graphitic foams~\cite{Li_GMR_graphitic_foam} and twisted graphene spirals~\cite{MR_spirals}. The latter fact encourages us to expect large off-diagonal thermopower in CVD multilayer graphene.

Here, we report on large Nernst effect in CVD graphene multilayers under ambient conditions in the field of permanent magnet. Localized temperature gradient is induced by heating the film with mid-infrared laser focused to the spot with diameter order of tens of microns. The emerging photo-thermoelectric voltages are measured in the rectangle-shaped sample upon illumination of its different sides. We find that the transverse (Nernst) voltage exceeds the longitudinal one (normal Seebeck effect) already at $B$ above 4 mT. The longitudinal voltage displays strong spatial fluctuations signifying the presence of electron and hole puddles in a non-uniform film. The transverse voltage has persistent sign, which is the indicator of electron-hole codirectional motion under simultaneous action of $B$ and $\nabla T$. 

\begin{figure}[ht]
    \includegraphics[width=1.0\linewidth]{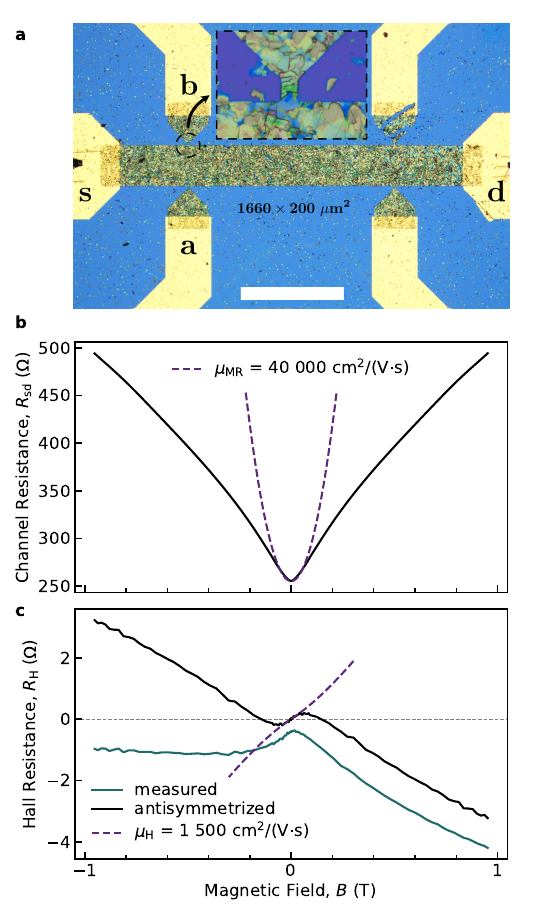}
    \caption{{\bf Device structure and basic characterization.} (a) Micro-photograph of the Hall-bar device based on a macroscopic graphitic film. Scale bar is 500 $\mu$m. Inset shows the magnified view of an individual Hall probe. (b) Measured room-temperature magnetoresistance of the film (solid) and its low-field fit quadratic fit ${\rm MR} = (\mu B)^2$ (dashed) for the mobility $\mu \approx 39\times 10^3$ cm$^2$/(V s) (c) As measured (green) and anti-symmetrized (black) dependences of Hall resistance on the magnetic field along with the linear fit for Hall mobility $\mu_H \approx 1.5\times 10^3$ cm$^2$/(V s)}
    \label{fig1}
\end{figure}

The sample under study is shown in Fig.~\ref{fig1}a. It represents a Hall-bar shaped CVD graphene film with dimensions $L\times W = 1660$ $\mu$m $\times 200$ $\mu$m. Source and drain contacts are made to short sides, and four Hall contacts are made symmetrically to longer sides. The film growth protocol can be found in Refs.~\cite{Kononenko2022,Brzhezinskaya2021} and in Supplementary section I. Raman spectroscopy~\cite{Kononenko2022} indicates the presence of random twists between graphene layers constituting the film, concordant with reports of interlayer twist by other groups~\cite{Synt_Garlow_TurbostraticMLG_Raman,Synt_Wu_Raman_TMLG,Synt_Yang_Turbostratcic_Self_Heating}. Analysis of sample micro-photographs (inset of Fig.~\ref{fig1}a) reveals the presence of strong thickness variations between 20 nm and 100 nm, each grain having the characteristic diameter of $\sim 10$ $\mu$m

We start the film characterization with magnetoresistance (MR) and Hall effect in the magnetic fields up to 1 T controlled by an external electromagnet. All measurements were performed at room temperature and atmospheric pressure. The resistance measured between source and drain contacts $R_{sd}(B)$ is shown in Fig.~\ref{fig1} (b). The MR attains a very large value $R(B=1 \, {\rm T})/R(B=0) - 1 = 100$ \% which we attribute to high mobility of electrons in graphene and twist-induced interlayer decoupling. The normal quadratic-in-$B$ region of MR is very narrow and is limited to the fields $B\lesssim 50$ mT; at all other fields the MR is almost linear. Quadratic fits of initial part of ${\rm MR}(B)$ curves yield the carrier mobilities of $(3...5)\times 10^4$ cm$^2$/V s, the definite value depends on the range of fields used for the fitting (see Supplementary section II). The measured Hall voltage, shown in Fig.~\ref{fig1} (c), is highly asymmetric in $B$ and resembles, in many details, the measured ${\rm MR}(B)$--curve. It implies that the residual Ohmic voltage drop between Hall contacts is comparable to the (weak) Hall voltage. Anti-symmetrization of transverse voltage with respect to $B$ allows us to extract pure Hall resistance $R_{\rm H}(B)$, which is shown in Fig.~\ref{fig1} (c) with blue line. The pure Hall resistance changes sign at $B\approx 50$ mT, while the linear fits of $R_{\rm H}(B)$ yield low apparent mobilities $\mu_{\rm H}\approx 1.5\times10^3$ cm$^2$/V s.

\begin{figure*}[ht!]
    \includegraphics[width=1.0\linewidth]{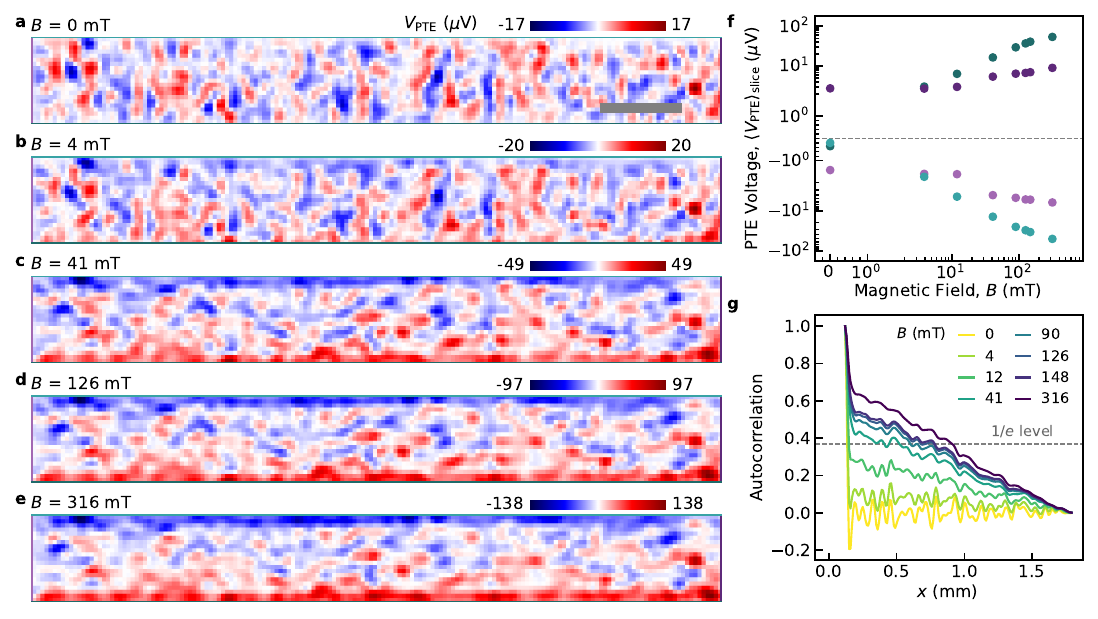}
    \caption{{\bf Thermoelectric voltage in graphitic film induced by selective laser heating.} (a-e) Maps of the measured PTE voltage between source and drain contacts for magnetic fields from zero to 315 mT, the maximum field available with the permanent centimeter-sized magnet. Scale bar is 200~$\mu$m. (f) Field dependence of spatially-averaged PTE voltage $\langle V_{\rm PTE}\rangle_{\rm slice}$, the averaging is performed over four slices of the full map marked by colored lines. Violet points correspond to the longitudinal excitation at the source and drain contacts; azure lines correspond to the transverse excitation at the  top and bottom edges of the sample. (g) Autocorrelation functions of PTE voltage at different laser beam positions $\langle V_{\rm PTE}({\bf r})V_{\rm PTE}({\bf r}+\Delta {\bf r})\rangle$ at various magnetic fields as a function of distance $|\Delta{\bf r}|$}
    \label{fig2}
\end{figure*}

Both linear magnetoresistance and sign-changing Hall resistance can be explained by the strong film non-uniformity, where the role of charge accumulation at the boundaries of grains with different conductivity is determinitial for magnetotransport. Linear magnetoresistance in such films was demonstrated with numerical simulations~\cite{EMT_magnetoresistance_2}, analytical duality arguments and semi-analytical effective medium theory (EMT)~\cite{EMT_magnetoresistance_1}. Sign-changing Hall resistance also appears at the level of EMT~\cite{EMT_hall_effect}. We note that exactly linear MR in such theories appears at equal fillings of the film by electron- and hole-doped regions, otherwise MR tends to saturation. Further measurements bring extra evidence for such $50/50$ electron-hole filling. 

Our measurements of thermopower rely on spatially-resolved heating of the sample by the focused infrared beam, and readout of the emerging photo-thermoelectric (PTE) voltage (see Supplementary section II). A similar technique, yet with sub-wavelength resolution, was recently used to study local thermopower of twisted graphene~\cite{Stepanov_TTG_Thermopower,Basov_LocalThermopower}. Our sample is mounted on the motorized stage, a special pocket behind it enables the placement of permanent magnets of different size. The precise value of $B$ is determined by measurement of the instant $R_{\rm sd}(B)$ and use of the calibration curve of Fig.~\ref{fig1} (b). The film is irradiated by a quantum cascade laser with wavelength $\lambda_0 = 8.1$ $\mu$m and power $P_0 \approx 4$ mW focused to the spot with size $\sigma \approx 20$ $\mu$m~\footnote{The size of the focused beam was estimated in the independent series of experiments, where the photovoltage maps of miniature photodetectors were recorded and fitted with Gaussian function $V_{\rm ph}(\bf r) \propto \exp\{-|{\bf r}|^2/2\sigma^2\}$. Best fits yield $\sigma \approx 20$ $\mu$m, thus the spatial distribution of the lase power density can be presented as $p({\bf r}) = P_0 (2\pi\sigma^2)^{-1} \exp\{-|{\bf r}|^2/2\sigma^2\}$ }. Absorption of laser radiation heats the sample to the temperature $T({\bf r})$ different from the base temperature $T_0$. If the heating is non-uniform and/or the thermopower $S({\bf r})$ is position-dependent, finite PTE voltage arises between contacts. In a prototypical case of rectangular uniform film with metallic source and drain, voltages of opposite signs would appear upon illumination of opposite contacts~\cite{Xia_photocurrent_imaging,Gabor_junction_PTE}. Illumination in the bulk of the film would result in negligible PTE voltage. Indeed, PTE requires large temperature difference between the two junctions and is maximized when one junction is selectively illuminated while the other retains dark. Illumination far from the metal results in low temperature difference between contacts.

The mapping of PTE voltage reveals a pattern strikingly different from that in uniform films, particularly, single layer graphene~\cite{Gabor_junction_PTE,Xia_photocurrent_imaging}. This pattern at $B=0$ is shown in Fig.~\ref{fig2} (a). The PTE voltage displays strong fluctuations both in sign and magnitude with characteristic autocorrelation length $l_{\rm AC} =5$ $\mu$m. Fluctuation amplitude does not depend on the distance between laser beam and the source/drain contact with two exceptions. In extreme proximity to the source contact, the PTE voltage is mostly negative, while in the vicinity of the drain contact, it is mostly positive. These observations can be explained by random electron-hole doping of the graphitic film in the bulk, and fixed metal-induced doping at the contacts. Each time the laser beam crosses and heats the random $p-n$ junction, the magnitude of PTE voltage is maximized, while when the laser illuminates the locally-uniform area, the PTE voltage is close to zero. Similar random PTE voltage patterns were observed with near-field photocurrent microscopy of CVD monolayer graphene~\cite{Woessner_nanoscopy}, yet with shorter autocorrelation length $l_{\rm AC}\sim200$ nm. In the proximity of metal leads, the metal-graphite work function difference results in large film doping, which overwhelms the local fluctuations. This results in spatially permanent photovoltage upon illumination of the contact.

The maps of photo-thermoelectric voltage change completely upon application of a weak magnetic field. Examples of maps at $B=4...316$ mT are shown in Fig.~\ref{fig2} (b-e). Full set of recorded maps is presented in Supplementary section III, while measurements of extra sample of smaller size are presented in Supplementary section IV. At finite $B$, random puddles of negative photovoltage coalesce into a sea at the top sample edge, with several positive photovoltage puddles forming isolated islands. The situation at the bottom edge is inverted and corresponds to the positive sea with negative islands. The average magnitude of PTE voltage at top and bottom edges increases almost linearly with the applied field, as shown in Fig.~\ref{fig2} (f). Fluctuations of voltage sign disappear completely at the edges above $B\approx 10...20$ mT. The effect can be quantified by a rapid increase in voltage correlation length with increase in $B$, illustrated in Fig.~\ref{fig2} g. Importantly, the fluctuations of voltage sign persist in the middle of the sample up to the largest fields.

Emergence of PTE voltage upon illumination of top and bottom sample edges is precisely the photo-Nernst effect. Once the laser spot is centered at the edge, non-zero average temperature gradient $\langle \partial_y T \rangle$ is induced in the sample, which results in transverse voltage $V^{(t)}_{\rm PTE} \propto S_{xy} \langle \partial_y T \rangle$. Electron and hole currents upon Nernst effect are co-directional. This fact manifests as smearing of electron-hole puddles in the photovoltage maps, and emergence of fixed-sign signal upon edge illumination. Once the laser spot moves away from the edge, the average temperature gradient $\langle \partial_y T \rangle$ rapidly decreases, and the Nernst effect surrenders to the sign-alternating diagonal Seebeck effect.

We proceed to estimate the magnitude of thermopower $S_{xx}$ and $S_{xy}(B)$ from our measurements. These are the central quantities of interest for the design of thermoelectric generators. We start with expression for source-drain PTE current as a weighted average of local currents~\cite{Levitov_Osng_SHockley_Ramo}:
\begin{equation}
\label{eq-LSSR}
    I_{\rm PTE} = \int{{\bf w}({\bf r}){\bf j}_{\rm PTE}({\bf r}) d^2{\bf r}},
\end{equation}
where ${\bf j}_{\rm PTE}({\bf r})=\hat{\alpha} \nabla T({\bf r})$ is the local current density, $\hat{\alpha}=\hat{S}/\rho$ is the  differential thermopower tensor, $\rho$ is the film resistivity, and ${\bf w}({\bf r})$ is the vector weight function with the dimension of inverse length. Physically, ${\bf w}({\bf r})$ is the electric field upon application of unit voltage drop between terminals. For measurement between wide parallel source and drain leads, ${\bf w}({\bf r}) = \{1/L,0\}$, and the current becomes proportional to the average temperature difference between sample edges. Eventually, we link these temperatures to the known laser power by solving the heat conduction equation in the Si/SiO$_2$ substrate (Supplementary section V). It is convenient to consider separately the cases of longitudinal and transverse excitation of temperature gradients, the corresponding voltages denoted by $V^{(l)}_{\rm PTE}$ and $V^{(t)}_{\rm PTE}$. In the former case, the beam is asymmetric with respect to the $y$-axis, the corresponding voltage maximized upon illumination of either source or drain junctions. In the latter case, the beam is asymmetric with respect to the $x$-axis, the corresponding voltage is maximized upon illumination of top and bottom edges in the magnetic field. We have obtained the following expressions:
\begin{equation}
\label{eq-l-t-voltgaes}
\left( \begin{aligned}
 V_{\rm PTE}^{\left( l \right)} \\ 
 V_{\rm PTE}^{\left( l \right)} \\ 
\end{aligned} \right)
=\left( \begin{aligned}
{{C}_{l}}{{S}_{xx}} \\ 
{{C}_{t}}{{S}_{xy}} \\ 
\end{aligned} \right)\times \frac{{{P}_{0}}}{\sigma {{\kappa }_{\text{Si}}}},
\end{equation}
where $P_0/(\sigma {\kappa }_{\text{Si}})\approx 1.3$ K is the characteristic increase in film temperature upon absorption of power $P_0$ in a spot of size $\sigma$, $C_l$ and $C_t$ are the geometric factors depending on channel dimensions, beam width, thickness and thermal conductivity of SiO$_2$ layer. Detailed evaluation in Supplementary section V shows that $C_l \approx C_t \approx 0.14$. Before actual estimates, we note that the developed theory deals with uniform channel conductivity and thermopower, therefore, it predicts only fluctuation-averaged (effective-medium) values of $S_{xx}$ and $S_{xy}$.

\begin{figure}[ht]
    \includegraphics[width=1.0\linewidth]{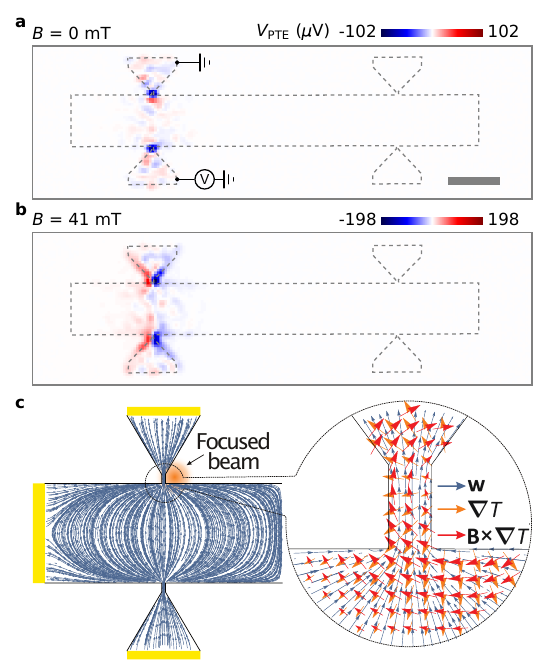}
    \caption{{\bf Enhancement of transverse photovoltage by contact pair selection.} (a,b) Maps of local PTE voltage measured between Hall contacts at $B=0$ (a) and $B=41$ mT (b). (c) Schematic illustrating the origins of enhanced off-diagonal PTE at the Hall contacts: flow lines of the virtual current ${\bf w}({\bf r})$ (blue), vector field of the temperature gradient $\nabla T$ (orange), and vector field of the Nernst current $[{\bf B},\nabla T]$ (red). Laser excitation adjacent to the apex of Hall probe leads to strong spatial overlap of ${\bf w}({\bf r})$ and ${\bf j}_{\rm PTE}\propto[{\bf B},\nabla T]$}
    \label{fig3}
\end{figure}

The fluctuation-averaged PTE voltages computed from the complete maps are shown in Fig.~\ref{fig2} (f) separately for the excitation of left/right sample edges (longitudinal case) and top/bottom sample edges (transverse case). In the absence of magnetic field, $V^{(l)}_{\rm PTE}$ at the source and drain junctions is about 2 $\mu$V, which translates into thermopower $S_{xx} \sim 10$ $\mu$V/K. This thermopower characterizes the metal-doped graphitic film in the vicinity of the junction. The sign of the PTE voltage corresponds to the electron doping of the film. Average PTE voltage at $B=0$ upon illumination of top and bottom edges is an order of magnitude smaller, such that the residual average $S_{xx}$ in the bulk is estimated below 1 $\mu$V/K.

Already in a very weak magnetic field $B=4$ mT, the spatially-averaged transverse thermoelectric voltage becomes equal to the longitudinal voltage at the contact. Thus, we estimate $S_{xy} \sim 10$ $\mu$V/K at this particular magnetic field. At the largest available field $B = 315$ mT, the average transverse voltage reaches 50 $\mu V$, while $S_{xy}$ is as large as 270 $\mu$V/K. The Nernst coefficient ${\rm N} = S_{xy}/B$ in weak (strong) magnetic fields is estimated as 2500 $\mu$V/(K T) [800 $\mu$V/(K T)]. This value is well above that for compensated InSb, the prototypical high-mobility semiconductor, where ${\rm N}\approx 100$ $\mu$ V/(K T) was measured~\cite{InSb_Nernst,InSb_Nernst_1}, and above that for trilayer graphene, where ${\rm N}\approx 110$ $\mu$ V/(K T) was predicted from modeling~\cite{ABA_graphene_nernst}.

We conclude our presentation of Nernst effect in graphitic films with a peculiar approach to enhance $V_{\rm PTE}$ at given temperature gradient. This is achieved by proper relative positioning of laser spot and voltage terminals. In Fig.~\ref{fig3} (a,b), where we present the map of PTE voltage between two Hall probes. These probes are triangle-shaped with narrow apex impinging into the sample. At zero field, the signal again represents spatial voltage fluctuations, which amplitude is enhanced near the wedges (Fig.~\ref{fig3} a). At $B=41$ mT (Fig.~\ref{fig3} b), the signal at the contact point becomes large and reaches $V_{\rm PTE,\max}\approx 200$ $\mu$V. Its sign changes as the laser beam crosses the wedge. For comparison, the maximum signal for source-drain voltage measurement at the same field is only 45 $\mu$V.

To explain this unusual enhancement, we turn again to the rule (\ref{eq-LSSR}) relating local and global thermoelectric currents. Once wedge-shaped Hall contacts are used for the measurement, the sampling field ${\bf w}({\bf r})$ is amplified and concentrated at contact constrictions. If the local current density ${\bf j}_{\rm PTE}\propto {\rm N} [{\bf B},\nabla T]$ has a strong spatial overlap with ${\bf w}({\bf r})$, the terminal current is largely enhanced. Simultaneous mapping of ${\bf w}({\bf r})$ and ${\bf j}_{\rm PTE}$ (Fig.~\ref{fig3} c) shows that maximization of the scalar product ${\bf w}({\bf r}) {\bf j}_{\rm PTE}({\bf r})$ occurs right when the laser beam is adjacent to the either side of contact apex.

Concluding our presentation, we discuss several limitations of the experimental technique and theoretical model for the estimate of thermopower tensor. First, the temperature $T({\bf r})$ of the film is not measured but estimated from heat conduction equation. Experimental scheme with selective Joule-heating and on-chip thermometer~\cite{Kim_thermopower} would be free of this shortcoming. Estimate of maximum laser-induced temperature rise $T_{\max}-T_0$ is relies on assumption of complete infrared absorption by the graphitic layer, which is realistic given the large number of atomic planes $N > 200$ in our film. Independent measurements of infrared absorption and transmission may improve this estimate~\cite{Gaskill_NL_Infrared_C_face}. Laser excitation may lead to other types of photovoltage different from thermoelectricity, particularly, photovoltaic effect at random $p-n$ junctions. However, photovoltaic effect is suppressed due to the rapid electron-hole Auger recombination in graphene~\cite{Alymov2018a} and its multilayers~\cite{AR_multilayers}.

Independent of these estimates is the experimental fact of comparable transverse thermoelectric voltage at $B^*\approx 4$ mT and the longitudinal voltage at zero magnetic field in the vicinity of metal contact. It shows that weak magnetic fields are no less promising for induction of thermoelectricity in graphite than metal-induced doping. The practical thermoelectric harvesting can be (speculatively) achieved in composite media comprised of graphitic films and nanoparticles of ferromagnetic metals, such as iron or nickel.

The CVD multilayer films in our study possessed high intrinsic mobility $\mu_{\rm MR}=(3...5)\times 10^4$ cm$^2$/(V s), which promoted large Nernst effect. The magnitude of $\mu_{\rm MR}$ in our films exceeds the characteristic value for graphite $\mu_{\rm gr}\approx 10^4$ cm$^2$/(V s)~\cite{mcclure1958analysis}, presumably due to interlayer decoupling induced by twist. Similar decoupling and mobility enhancement was evidenced in pyrolythic graphene on SiC~\cite{MLG_on_SiC}. Whether large Nernst effect would persist in non-twisted CVD graphene multilayers remains an open question, both from experimental and theoretical perspectives. While extensive theoretical literature exists on magnetoresistance~\cite{EMT_magnetoresistance_2,EMT_magnetoresistance_1}, Hall effect~\cite{EMT_hall_effect} and Seebeck effect~\cite{balagurov1986theory} in non-uniform thin films, the studies of Nernst effect are limited to the low-field limit only~\cite{Fishchuk_Nernst_Disordered}. Derivation of a theoretical link between $\mu$ and ${\rm N}$ in disordered thin films can shed light on precise observation conditions of large Nernst effect.

{\it Acknowledgments.} The authors thank Alexander I. Chernov for providing an electromagnet for measurements, and Alexander D. Morozov for 3d printing of the sample holder.

{\it Funding.} The work was supported by the Ministry of Science and Higher Education of the Russian Federation, agreement \# 075-15-2025-608.

{\it Data availability.} The data that supports the findings of this study are available from the corresponding author upon reasonable request.

\bibliography{references}

\begin{thebibliography}{46}%
\makeatletter
\providecommand \@ifxundefined [1]{%
 \@ifx{#1\undefined}
}%
\providecommand \@ifnum [1]{%
 \ifnum #1\expandafter \@firstoftwo
 \else \expandafter \@secondoftwo
 \fi
}%
\providecommand \@ifx [1]{%
 \ifx #1\expandafter \@firstoftwo
 \else \expandafter \@secondoftwo
 \fi
}%
\providecommand \natexlab [1]{#1}%
\providecommand \enquote  [1]{``#1''}%
\providecommand \bibnamefont  [1]{#1}%
\providecommand \bibfnamefont [1]{#1}%
\providecommand \citenamefont [1]{#1}%
\providecommand \href@noop [0]{\@secondoftwo}%
\providecommand \href [0]{\begingroup \@sanitize@url \@href}%
\providecommand \@href[1]{\@@startlink{#1}\@@href}%
\providecommand \@@href[1]{\endgroup#1\@@endlink}%
\providecommand \@sanitize@url [0]{\catcode `\\12\catcode `\$12\catcode `\&12\catcode `\#12\catcode `\^12\catcode `\_12\catcode `\%12\relax}%
\providecommand \@@startlink[1]{}%
\providecommand \@@endlink[0]{}%
\providecommand \url  [0]{\begingroup\@sanitize@url \@url }%
\providecommand \@url [1]{\endgroup\@href {#1}{\urlprefix }}%
\providecommand \urlprefix  [0]{URL }%
\providecommand \Eprint [0]{\href }%
\providecommand \doibase [0]{https://doi.org/}%
\providecommand \selectlanguage [0]{\@gobble}%
\providecommand \bibinfo  [0]{\@secondoftwo}%
\providecommand \bibfield  [0]{\@secondoftwo}%
\providecommand \translation [1]{[#1]}%
\providecommand \BibitemOpen [0]{}%
\providecommand \bibitemStop [0]{}%
\providecommand \bibitemNoStop [0]{.\EOS\space}%
\providecommand \EOS [0]{\spacefactor3000\relax}%
\providecommand \BibitemShut  [1]{\csname bibitem#1\endcsname}%
\let\auto@bib@innerbib\@empty
\bibitem [{\citenamefont {Shi}\ \emph {et~al.}(2025)\citenamefont {Shi}, \citenamefont {Li}, \citenamefont {Li},\ and\ \citenamefont {Chen}}]{TE_materials_review}%
  \BibitemOpen
  \bibfield  {author} {\bibinfo {author} {\bibfnamefont {X.-L.}\ \bibnamefont {Shi}}, \bibinfo {author} {\bibfnamefont {N.-H.}\ \bibnamefont {Li}}, \bibinfo {author} {\bibfnamefont {M.}~\bibnamefont {Li}},\ and\ \bibinfo {author} {\bibfnamefont {Z.-G.}\ \bibnamefont {Chen}},\ }\bibfield  {title} {\bibinfo {title} {Toward efficient thermoelectric materials and devices: Advances, challenges, and opportunities},\ }\href {https://doi.org/10.1021/acs.chemrev.5c00060} {\bibfield  {journal} {\bibinfo  {journal} {Chemical Reviews}\ }\textbf {\bibinfo {volume} {125}},\ \bibinfo {pages} {7525} (\bibinfo {year} {2025})}\BibitemShut {NoStop}%
\bibitem [{\citenamefont {Mulla}\ \emph {et~al.}(2023)\citenamefont {Mulla}, \citenamefont {White}, \citenamefont {Dunnill},\ and\ \citenamefont {Barron}}]{TE_Applications_graphene}%
  \BibitemOpen
  \bibfield  {author} {\bibinfo {author} {\bibfnamefont {R.}~\bibnamefont {Mulla}}, \bibinfo {author} {\bibfnamefont {A.~O.}\ \bibnamefont {White}}, \bibinfo {author} {\bibfnamefont {C.~W.}\ \bibnamefont {Dunnill}},\ and\ \bibinfo {author} {\bibfnamefont {A.~R.}\ \bibnamefont {Barron}},\ }\bibfield  {title} {\bibinfo {title} {The role of graphene in new thermoelectric materials},\ }\href {https://doi.org/10.1039/d3ya00085k} {\bibfield  {journal} {\bibinfo  {journal} {Energy Advances}\ }\textbf {\bibinfo {volume} {2}},\ \bibinfo {pages} {606} (\bibinfo {year} {2023})}\BibitemShut {NoStop}%
\bibitem [{\citenamefont {Slonczewski}\ and\ \citenamefont {Weiss}(1958)}]{Slonch_Weiss_bands}%
  \BibitemOpen
  \bibfield  {author} {\bibinfo {author} {\bibfnamefont {J.}~\bibnamefont {Slonczewski}}\ and\ \bibinfo {author} {\bibfnamefont {P.}~\bibnamefont {Weiss}},\ }\bibfield  {title} {\bibinfo {title} {Band structure of graphite},\ }\href {https://doi.org/10.1103/PhysRev.109.272} {\bibfield  {journal} {\bibinfo  {journal} {Physical review}\ }\textbf {\bibinfo {volume} {109}},\ \bibinfo {pages} {272} (\bibinfo {year} {1958})}\BibitemShut {NoStop}%
\bibitem [{\citenamefont {McClure}(1958)}]{mcclure1958analysis}%
  \BibitemOpen
  \bibfield  {author} {\bibinfo {author} {\bibfnamefont {J.}~\bibnamefont {McClure}},\ }\bibfield  {title} {\bibinfo {title} {Analysis of multicarrier galvanomagnetic data for graphite},\ }\href {https://doi.org/10.1103/PhysRev.112.715} {\bibfield  {journal} {\bibinfo  {journal} {Physical Review}\ }\textbf {\bibinfo {volume} {112}},\ \bibinfo {pages} {715} (\bibinfo {year} {1958})}\BibitemShut {NoStop}%
\bibitem [{\citenamefont {Rositawati}\ \emph {et~al.}(2024)\citenamefont {Rositawati}, \citenamefont {Widianto}, \citenamefont {Suprapto}, \citenamefont {Sujitno}, \citenamefont {Absor}, \citenamefont {Sholihun}, \citenamefont {Triyana},\ and\ \citenamefont {Santoso}}]{TE_enhancement_metal_induced}%
  \BibitemOpen
  \bibfield  {author} {\bibinfo {author} {\bibfnamefont {D.~N.}\ \bibnamefont {Rositawati}}, \bibinfo {author} {\bibfnamefont {E.}~\bibnamefont {Widianto}}, \bibinfo {author} {\bibnamefont {Suprapto}}, \bibinfo {author} {\bibfnamefont {T.}~\bibnamefont {Sujitno}}, \bibinfo {author} {\bibfnamefont {M.~A.~U.}\ \bibnamefont {Absor}}, \bibinfo {author} {\bibnamefont {Sholihun}}, \bibinfo {author} {\bibfnamefont {K.}~\bibnamefont {Triyana}},\ and\ \bibinfo {author} {\bibfnamefont {I.}~\bibnamefont {Santoso}},\ }\bibfield  {title} {\bibinfo {title} {Enhancing thermoelectric properties of multilayer graphene with au deposition},\ }\href {https://doi.org/https://doi.org/10.1016/j.matchemphys.2024.129295} {\bibfield  {journal} {\bibinfo  {journal} {Materials Chemistry and Physics}\ }\textbf {\bibinfo {volume} {319}},\ \bibinfo {pages} {129295} (\bibinfo {year} {2024})}\BibitemShut {NoStop}%
\bibitem [{\citenamefont {Duan}\ \emph {et~al.}(2024)\citenamefont {Duan}, \citenamefont {Meng}, \citenamefont {Wu}, \citenamefont {Yang}, \citenamefont {Zhong}, \citenamefont {Ao}, \citenamefont {Yao}, \citenamefont {Fang},\ and\ \citenamefont {Huang}}]{Fe_doping_graphite}%
  \BibitemOpen
  \bibfield  {author} {\bibinfo {author} {\bibfnamefont {S.}~\bibnamefont {Duan}}, \bibinfo {author} {\bibfnamefont {K.}~\bibnamefont {Meng}}, \bibinfo {author} {\bibfnamefont {X.}~\bibnamefont {Wu}}, \bibinfo {author} {\bibfnamefont {M.}~\bibnamefont {Yang}}, \bibinfo {author} {\bibfnamefont {M.}~\bibnamefont {Zhong}}, \bibinfo {author} {\bibfnamefont {W.}~\bibnamefont {Ao}}, \bibinfo {author} {\bibfnamefont {Y.}~\bibnamefont {Yao}}, \bibinfo {author} {\bibfnamefont {M.}~\bibnamefont {Fang}},\ and\ \bibinfo {author} {\bibfnamefont {Z.}~\bibnamefont {Huang}},\ }\bibfield  {title} {\bibinfo {title} {Preparation and properties of graphite-based “light–heat–electricity” conversion materials},\ }\href {https://doi.org/10.1063/5.0239344} {\bibfield  {journal} {\bibinfo  {journal} {Applied Physics Letters}\ }\textbf {\bibinfo {volume} {125}},\ \bibinfo {pages} {244101} (\bibinfo {year} {2024})}\BibitemShut {NoStop}%
\bibitem [{\citenamefont {Hwang}\ \emph {et~al.}(2023)\citenamefont {Hwang}, \citenamefont {Kim}, \citenamefont {Lee},\ and\ \citenamefont {Lee}}]{Gradient_doping_graphite}%
  \BibitemOpen
  \bibfield  {author} {\bibinfo {author} {\bibfnamefont {H.~J.}\ \bibnamefont {Hwang}}, \bibinfo {author} {\bibfnamefont {S.-Y.}\ \bibnamefont {Kim}}, \bibinfo {author} {\bibfnamefont {S.~K.}\ \bibnamefont {Lee}},\ and\ \bibinfo {author} {\bibfnamefont {B.~H.}\ \bibnamefont {Lee}},\ }\bibfield  {title} {\bibinfo {title} {Large scale graphene thermoelectric device with high power factor using gradient doping profile},\ }\href {https://doi.org/https://doi.org/10.1016/j.carbon.2022.09.048} {\bibfield  {journal} {\bibinfo  {journal} {Carbon}\ }\textbf {\bibinfo {volume} {201}},\ \bibinfo {pages} {467} (\bibinfo {year} {2023})}\BibitemShut {NoStop}%
\bibitem [{\citenamefont {Matsumoto}\ \emph {et~al.}(2009)\citenamefont {Matsumoto}, \citenamefont {Hoshina},\ and\ \citenamefont {Akuzawa}}]{Intercalation_doping_graphite}%
  \BibitemOpen
  \bibfield  {author} {\bibinfo {author} {\bibfnamefont {R.}~\bibnamefont {Matsumoto}}, \bibinfo {author} {\bibfnamefont {Y.}~\bibnamefont {Hoshina}},\ and\ \bibinfo {author} {\bibfnamefont {N.}~\bibnamefont {Akuzawa}},\ }\bibfield  {title} {\bibinfo {title} {Thermoelectric properties and electrical transport of graphite intercalation compounds},\ }\href {https://doi.org/10.2320/matertrans.E-M2009813} {\bibfield  {journal} {\bibinfo  {journal} {Materials Transactions}\ }\textbf {\bibinfo {volume} {50}},\ \bibinfo {pages} {1607} (\bibinfo {year} {2009})}\BibitemShut {NoStop}%
\bibitem [{\citenamefont {Rahimi}\ \emph {et~al.}(2026)\citenamefont {Rahimi}, \citenamefont {Lubertino}, \citenamefont {Bellelli}, \citenamefont {Chen}, \citenamefont {Lafarge}, \citenamefont {Barraud}, \citenamefont {Mallet}, \citenamefont {Martin}, \citenamefont {Chaste}, \citenamefont {Fournier},\ and\ \citenamefont {Della~Rocca}}]{TE_enhancement_nanomeshing}%
  \BibitemOpen
  \bibfield  {author} {\bibinfo {author} {\bibfnamefont {M.}~\bibnamefont {Rahimi}}, \bibinfo {author} {\bibfnamefont {N.}~\bibnamefont {Lubertino}}, \bibinfo {author} {\bibfnamefont {R.}~\bibnamefont {Bellelli}}, \bibinfo {author} {\bibfnamefont {L.}~\bibnamefont {Chen}}, \bibinfo {author} {\bibfnamefont {P.}~\bibnamefont {Lafarge}}, \bibinfo {author} {\bibfnamefont {C.}~\bibnamefont {Barraud}}, \bibinfo {author} {\bibfnamefont {F.}~\bibnamefont {Mallet}}, \bibinfo {author} {\bibfnamefont {P.}~\bibnamefont {Martin}}, \bibinfo {author} {\bibfnamefont {J.}~\bibnamefont {Chaste}}, \bibinfo {author} {\bibfnamefont {D.}~\bibnamefont {Fournier}},\ and\ \bibinfo {author} {\bibfnamefont {M.}~\bibnamefont {Della~Rocca}},\ }\bibfield  {title} {\bibinfo {title} {Decoupling thermoelectric coefficients of multilayer graphene by nanomeshing},\ }\href {https://doi.org/10.1103/9jsm-f2rx} {\bibfield  {journal} {\bibinfo  {journal} {Phys. Rev. Appl.}\ }\textbf {\bibinfo {volume} {25}},\ \bibinfo {pages} {024039} (\bibinfo
  {year} {2026})}\BibitemShut {NoStop}%
\bibitem [{\citenamefont {Zuev}\ \emph {et~al.}(2009)\citenamefont {Zuev}, \citenamefont {Chang},\ and\ \citenamefont {Kim}}]{Kim_thermopower}%
  \BibitemOpen
  \bibfield  {author} {\bibinfo {author} {\bibfnamefont {Y.~M.}\ \bibnamefont {Zuev}}, \bibinfo {author} {\bibfnamefont {W.}~\bibnamefont {Chang}},\ and\ \bibinfo {author} {\bibfnamefont {P.}~\bibnamefont {Kim}},\ }\bibfield  {title} {\bibinfo {title} {{Thermoelectric and Magnetothermoelectric Transport Measurements of Graphene}},\ }\href {https://doi.org/10.1103/PhysRevLett.102.096807} {\bibfield  {journal} {\bibinfo  {journal} {Physical Review Letters}\ }\textbf {\bibinfo {volume} {102}},\ \bibinfo {pages} {096807} (\bibinfo {year} {2009})}\BibitemShut {NoStop}%
\bibitem [{\citenamefont {Negishi}\ \emph {et~al.}(2020)\citenamefont {Negishi}, \citenamefont {Wei}, \citenamefont {Yao}, \citenamefont {Ogawa}, \citenamefont {Akabori}, \citenamefont {Kanai}, \citenamefont {Matsumoto}, \citenamefont {Taniyasu},\ and\ \citenamefont {Kobayashi}}]{Negishi_Turbostraticity_on_mobility}%
  \BibitemOpen
  \bibfield  {author} {\bibinfo {author} {\bibfnamefont {R.}~\bibnamefont {Negishi}}, \bibinfo {author} {\bibfnamefont {C.}~\bibnamefont {Wei}}, \bibinfo {author} {\bibfnamefont {Y.}~\bibnamefont {Yao}}, \bibinfo {author} {\bibfnamefont {Y.}~\bibnamefont {Ogawa}}, \bibinfo {author} {\bibfnamefont {M.}~\bibnamefont {Akabori}}, \bibinfo {author} {\bibfnamefont {Y.}~\bibnamefont {Kanai}}, \bibinfo {author} {\bibfnamefont {K.}~\bibnamefont {Matsumoto}}, \bibinfo {author} {\bibfnamefont {Y.}~\bibnamefont {Taniyasu}},\ and\ \bibinfo {author} {\bibfnamefont {Y.}~\bibnamefont {Kobayashi}},\ }\bibfield  {title} {\bibinfo {title} {{Turbostratic Stacking Effect in Multilayer Graphene on the Electrical Transport Properties}},\ }\href {https://doi.org/10.1002/pssb.201900437} {\bibfield  {journal} {\bibinfo  {journal} {physica status solidi (b)}\ }\textbf {\bibinfo {volume} {257}},\ \bibinfo {pages} {1} (\bibinfo {year} {2020})}\BibitemShut {NoStop}%
\bibitem [{\citenamefont {Paul}\ \emph {et~al.}(2022)\citenamefont {Paul}, \citenamefont {Ghosh}, \citenamefont {Chakraborty}, \citenamefont {Roy}, \citenamefont {Dutta}, \citenamefont {Watanabe}, \citenamefont {Taniguchi}, \citenamefont {Panda}, \citenamefont {Agarwala}, \citenamefont {Mukerjee} \emph {et~al.}}]{Interaction_thermopower}%
  \BibitemOpen
  \bibfield  {author} {\bibinfo {author} {\bibfnamefont {A.~K.}\ \bibnamefont {Paul}}, \bibinfo {author} {\bibfnamefont {A.}~\bibnamefont {Ghosh}}, \bibinfo {author} {\bibfnamefont {S.}~\bibnamefont {Chakraborty}}, \bibinfo {author} {\bibfnamefont {U.}~\bibnamefont {Roy}}, \bibinfo {author} {\bibfnamefont {R.}~\bibnamefont {Dutta}}, \bibinfo {author} {\bibfnamefont {K.}~\bibnamefont {Watanabe}}, \bibinfo {author} {\bibfnamefont {T.}~\bibnamefont {Taniguchi}}, \bibinfo {author} {\bibfnamefont {A.}~\bibnamefont {Panda}}, \bibinfo {author} {\bibfnamefont {A.}~\bibnamefont {Agarwala}}, \bibinfo {author} {\bibfnamefont {S.}~\bibnamefont {Mukerjee}}, \emph {et~al.},\ }\bibfield  {title} {\bibinfo {title} {Interaction-driven giant thermopower in magic-angle twisted bilayer graphene},\ }\href {https://doi.org/10.1038/s41567-022-01574-3} {\bibfield  {journal} {\bibinfo  {journal} {Nature Physics}\ }\textbf {\bibinfo {volume} {18}},\ \bibinfo {pages} {691} (\bibinfo {year} {2022})}\BibitemShut {NoStop}%
\bibitem [{\citenamefont {Zarenia}\ \emph {et~al.}(2019)\citenamefont {Zarenia}, \citenamefont {Principi},\ and\ \citenamefont {Vignale}}]{Zarenia_thermopower}%
  \BibitemOpen
  \bibfield  {author} {\bibinfo {author} {\bibfnamefont {M.}~\bibnamefont {Zarenia}}, \bibinfo {author} {\bibfnamefont {A.}~\bibnamefont {Principi}},\ and\ \bibinfo {author} {\bibfnamefont {G.}~\bibnamefont {Vignale}},\ }\bibfield  {title} {\bibinfo {title} {{Disorder-enabled hydrodynamics of charge and heat transport in monolayer graphene}},\ }\href {https://doi.org/10.1088/2053-1583/ab1ad9} {\bibfield  {journal} {\bibinfo  {journal} {2D Materials}\ }\textbf {\bibinfo {volume} {6}},\ \bibinfo {pages} {035024} (\bibinfo {year} {2019})}\BibitemShut {NoStop}%
\bibitem [{\citenamefont {Zhu}\ \emph {et~al.}(2010)\citenamefont {Zhu}, \citenamefont {Yang}, \citenamefont {Fauqu{\'{e}}}, \citenamefont {Kopelevich},\ and\ \citenamefont {Behnia}}]{Nernst_HOPG}%
  \BibitemOpen
  \bibfield  {author} {\bibinfo {author} {\bibfnamefont {Z.}~\bibnamefont {Zhu}}, \bibinfo {author} {\bibfnamefont {H.}~\bibnamefont {Yang}}, \bibinfo {author} {\bibfnamefont {B.}~\bibnamefont {Fauqu{\'{e}}}}, \bibinfo {author} {\bibfnamefont {Y.}~\bibnamefont {Kopelevich}},\ and\ \bibinfo {author} {\bibfnamefont {K.}~\bibnamefont {Behnia}},\ }\bibfield  {title} {\bibinfo {title} {{Nernst effect and dimensionality in the quantumlimit}},\ }\href {https://doi.org/10.1038/nphys1437} {\bibfield  {journal} {\bibinfo  {journal} {Nature Physics}\ }\textbf {\bibinfo {volume} {6}},\ \bibinfo {pages} {26} (\bibinfo {year} {2010})}\BibitemShut {NoStop}%
\bibitem [{\citenamefont {Cao}\ \emph {et~al.}(2016)\citenamefont {Cao}, \citenamefont {Aivazian}, \citenamefont {Fei}, \citenamefont {Ross}, \citenamefont {Cobden},\ and\ \citenamefont {Xu}}]{Nernst_Graphene}%
  \BibitemOpen
  \bibfield  {author} {\bibinfo {author} {\bibfnamefont {H.}~\bibnamefont {Cao}}, \bibinfo {author} {\bibfnamefont {G.}~\bibnamefont {Aivazian}}, \bibinfo {author} {\bibfnamefont {Z.}~\bibnamefont {Fei}}, \bibinfo {author} {\bibfnamefont {J.}~\bibnamefont {Ross}}, \bibinfo {author} {\bibfnamefont {D.~H.}\ \bibnamefont {Cobden}},\ and\ \bibinfo {author} {\bibfnamefont {X.}~\bibnamefont {Xu}},\ }\bibfield  {title} {\bibinfo {title} {{Photo-Nernst current in graphene}},\ }\href {https://doi.org/10.1038/nphys3549} {\bibfield  {journal} {\bibinfo  {journal} {Nature Physics}\ }\textbf {\bibinfo {volume} {12}},\ \bibinfo {pages} {236} (\bibinfo {year} {2016})}\BibitemShut {NoStop}%
\bibitem [{\citenamefont {Elesin}\ \emph {et~al.}(2026)\citenamefont {Elesin}, \citenamefont {Shilov}, \citenamefont {Jana}, \citenamefont {Mazurenko}, \citenamefont {Pantaleon}, \citenamefont {Kashchenko}, \citenamefont {Krivovichev}, \citenamefont {Dremov}, \citenamefont {Gayduchenko}, \citenamefont {Taniguchi}, \citenamefont {Watanabe}, \citenamefont {Wang}, \citenamefont {Novoselov}, \citenamefont {Svintsov}, \citenamefont {Goltsman}, \citenamefont {Titova},\ and\ \citenamefont {Bandurin}}]{Elesin_Nernst}%
  \BibitemOpen
  \bibfield  {author} {\bibinfo {author} {\bibfnamefont {L.}~\bibnamefont {Elesin}}, \bibinfo {author} {\bibfnamefont {A.~L.}\ \bibnamefont {Shilov}}, \bibinfo {author} {\bibfnamefont {S.}~\bibnamefont {Jana}}, \bibinfo {author} {\bibfnamefont {I.}~\bibnamefont {Mazurenko}}, \bibinfo {author} {\bibfnamefont {P.~A.}\ \bibnamefont {Pantaleon}}, \bibinfo {author} {\bibfnamefont {M.}~\bibnamefont {Kashchenko}}, \bibinfo {author} {\bibfnamefont {N.}~\bibnamefont {Krivovichev}}, \bibinfo {author} {\bibfnamefont {V.}~\bibnamefont {Dremov}}, \bibinfo {author} {\bibfnamefont {I.}~\bibnamefont {Gayduchenko}}, \bibinfo {author} {\bibfnamefont {T.}~\bibnamefont {Taniguchi}}, \bibinfo {author} {\bibfnamefont {K.}~\bibnamefont {Watanabe}}, \bibinfo {author} {\bibfnamefont {Y.}~\bibnamefont {Wang}}, \bibinfo {author} {\bibfnamefont {K.~S.}\ \bibnamefont {Novoselov}}, \bibinfo {author} {\bibfnamefont {D.~A.}\ \bibnamefont {Svintsov}}, \bibinfo {author} {\bibfnamefont {G.}~\bibnamefont {Goltsman}}, \bibinfo {author}
  {\bibfnamefont {E.~I.}\ \bibnamefont {Titova}},\ and\ \bibinfo {author} {\bibfnamefont {D.~A.}\ \bibnamefont {Bandurin}},\ }\bibfield  {title} {\bibinfo {title} {{Enhanced Terahertz Thermoelectricity Via Engineered Van Hove Singularities and Nernst Effect in Moir{\'{e}} Superlattices}},\ }\href {https://doi.org/10.1002/adfm.202528325} {\bibfield  {journal} {\bibinfo  {journal} {Advanced Functional Materials}\ }\textbf {\bibinfo {volume} {36}},\ \bibinfo {pages} {1} (\bibinfo {year} {2026})}\BibitemShut {NoStop}%
\bibitem [{\citenamefont {Roy}\ \emph {et~al.}(2026)\citenamefont {Roy}, \citenamefont {Roy}, \citenamefont {Ghorai}, \citenamefont {Mukherjee}, \citenamefont {Kumar}, \citenamefont {Watanabe}, \citenamefont {Taniguchi}, \citenamefont {Trivedi}, \citenamefont {Sensarma}, \citenamefont {Mukerjee},\ and\ \citenamefont {Das}}]{Nernst_TDBG}%
  \BibitemOpen
  \bibfield  {author} {\bibinfo {author} {\bibfnamefont {U.}~\bibnamefont {Roy}}, \bibinfo {author} {\bibfnamefont {M.}~\bibnamefont {Roy}}, \bibinfo {author} {\bibfnamefont {U.}~\bibnamefont {Ghorai}}, \bibinfo {author} {\bibfnamefont {A.}~\bibnamefont {Mukherjee}}, \bibinfo {author} {\bibfnamefont {R.}~\bibnamefont {Kumar}}, \bibinfo {author} {\bibfnamefont {K.}~\bibnamefont {Watanabe}}, \bibinfo {author} {\bibfnamefont {T.}~\bibnamefont {Taniguchi}}, \bibinfo {author} {\bibfnamefont {N.}~\bibnamefont {Trivedi}}, \bibinfo {author} {\bibfnamefont {R.}~\bibnamefont {Sensarma}}, \bibinfo {author} {\bibfnamefont {S.}~\bibnamefont {Mukerjee}},\ and\ \bibinfo {author} {\bibfnamefont {A.}~\bibnamefont {Das}},\ }\bibfield  {title} {\bibinfo {title} {{Van Hove singularity-driven giant Nernst signal in twisted double bilayer graphene}},\ }\href {http://arxiv.org/abs/2607.25359} {\bibfield  {journal} {\bibinfo  {journal} {arXiv preprint}\ } (\bibinfo {year} {2026})},\ \Eprint {https://arxiv.org/abs/2607.25359}
  {2607.25359} \BibitemShut {NoStop}%
\bibitem [{\citenamefont {Gong}\ \emph {et~al.}(2025)\citenamefont {Gong}, \citenamefont {Yang}, \citenamefont {Zhang}, \citenamefont {Pandey}, \citenamefont {Cui}, \citenamefont {Ruff}, \citenamefont {Horak}, \citenamefont {Karapetrova}, \citenamefont {Kim}, \citenamefont {Ryan}, \citenamefont {Hao}, \citenamefont {Zhang},\ and\ \citenamefont {Liu}}]{ANE_1}%
  \BibitemOpen
  \bibfield  {author} {\bibinfo {author} {\bibfnamefont {D.}~\bibnamefont {Gong}}, \bibinfo {author} {\bibfnamefont {J.}~\bibnamefont {Yang}}, \bibinfo {author} {\bibfnamefont {S.}~\bibnamefont {Zhang}}, \bibinfo {author} {\bibfnamefont {S.}~\bibnamefont {Pandey}}, \bibinfo {author} {\bibfnamefont {D.}~\bibnamefont {Cui}}, \bibinfo {author} {\bibfnamefont {J.~P.~C.}\ \bibnamefont {Ruff}}, \bibinfo {author} {\bibfnamefont {L.}~\bibnamefont {Horak}}, \bibinfo {author} {\bibfnamefont {E.}~\bibnamefont {Karapetrova}}, \bibinfo {author} {\bibfnamefont {J.-W.}\ \bibnamefont {Kim}}, \bibinfo {author} {\bibfnamefont {P.~J.}\ \bibnamefont {Ryan}}, \bibinfo {author} {\bibfnamefont {L.}~\bibnamefont {Hao}}, \bibinfo {author} {\bibfnamefont {Y.}~\bibnamefont {Zhang}},\ and\ \bibinfo {author} {\bibfnamefont {J.}~\bibnamefont {Liu}},\ }\bibfield  {title} {\bibinfo {title} {{Large asymmetric anomalous Nernst effect in the antiferromagnet SrIr0.8Sn0.2O3}},\ }\href {https://doi.org/10.1038/s41467-025-58020-0} {\bibfield
  {journal} {\bibinfo  {journal} {Nature Communications}\ }\textbf {\bibinfo {volume} {16}},\ \bibinfo {pages} {2888} (\bibinfo {year} {2025})}\BibitemShut {NoStop}%
\bibitem [{\citenamefont {Sakai}\ \emph {et~al.}(2018)\citenamefont {Sakai}, \citenamefont {Mizuta}, \citenamefont {Nugroho}, \citenamefont {Sihombing}, \citenamefont {Koretsune}, \citenamefont {Suzuki}, \citenamefont {Takemori}, \citenamefont {Ishii}, \citenamefont {Nishio-Hamane}, \citenamefont {Arita}, \citenamefont {Goswami},\ and\ \citenamefont {Nakatsuji}}]{ANE_2}%
  \BibitemOpen
  \bibfield  {author} {\bibinfo {author} {\bibfnamefont {A.}~\bibnamefont {Sakai}}, \bibinfo {author} {\bibfnamefont {Y.~P.}\ \bibnamefont {Mizuta}}, \bibinfo {author} {\bibfnamefont {A.~A.}\ \bibnamefont {Nugroho}}, \bibinfo {author} {\bibfnamefont {R.}~\bibnamefont {Sihombing}}, \bibinfo {author} {\bibfnamefont {T.}~\bibnamefont {Koretsune}}, \bibinfo {author} {\bibfnamefont {M.-T.}\ \bibnamefont {Suzuki}}, \bibinfo {author} {\bibfnamefont {N.}~\bibnamefont {Takemori}}, \bibinfo {author} {\bibfnamefont {R.}~\bibnamefont {Ishii}}, \bibinfo {author} {\bibfnamefont {D.}~\bibnamefont {Nishio-Hamane}}, \bibinfo {author} {\bibfnamefont {R.}~\bibnamefont {Arita}}, \bibinfo {author} {\bibfnamefont {P.}~\bibnamefont {Goswami}},\ and\ \bibinfo {author} {\bibfnamefont {S.}~\bibnamefont {Nakatsuji}},\ }\bibfield  {title} {\bibinfo {title} {{Giant anomalous Nernst effect and quantum-critical scaling in a ferromagnetic semimetal}},\ }\href {https://doi.org/10.1038/s41567-018-0225-6} {\bibfield  {journal} {\bibinfo
  {journal} {Nature Physics}\ }\textbf {\bibinfo {volume} {14}},\ \bibinfo {pages} {1119} (\bibinfo {year} {2018})}\BibitemShut {NoStop}%
\bibitem [{\citenamefont {Mizuguchi}\ and\ \citenamefont {Nakatsuji}(2019)}]{ANE_3}%
  \BibitemOpen
  \bibfield  {author} {\bibinfo {author} {\bibfnamefont {M.}~\bibnamefont {Mizuguchi}}\ and\ \bibinfo {author} {\bibfnamefont {S.}~\bibnamefont {Nakatsuji}},\ }\bibfield  {title} {\bibinfo {title} {{Energy-harvesting materials based on the anomalous Nernst effect}},\ }\href {https://doi.org/10.1080/14686996.2019.1585143} {\bibfield  {journal} {\bibinfo  {journal} {Science and Technology of Advanced Materials}\ }\textbf {\bibinfo {volume} {20}},\ \bibinfo {pages} {262} (\bibinfo {year} {2019})}\BibitemShut {NoStop}%
\bibitem [{\citenamefont {Goldsmid}\ \emph {et~al.}(1972)\citenamefont {Goldsmid}, \citenamefont {Knittel}, \citenamefont {Savvides},\ and\ \citenamefont {Uher}}]{InSb_Nernst}%
  \BibitemOpen
  \bibfield  {author} {\bibinfo {author} {\bibfnamefont {H.~J.}\ \bibnamefont {Goldsmid}}, \bibinfo {author} {\bibfnamefont {T.}~\bibnamefont {Knittel}}, \bibinfo {author} {\bibfnamefont {N.}~\bibnamefont {Savvides}},\ and\ \bibinfo {author} {\bibfnamefont {C.}~\bibnamefont {Uher}},\ }\bibfield  {title} {\bibinfo {title} {{Measurement of heat flow by means of the Nernst effect}},\ }\href {https://doi.org/10.1088/0022-3735/5/4/008} {\bibfield  {journal} {\bibinfo  {journal} {Journal of Physics E: Scientific Instruments}\ }\textbf {\bibinfo {volume} {5}},\ \bibinfo {pages} {313} (\bibinfo {year} {1972})}\BibitemShut {NoStop}%
\bibitem [{\citenamefont {Nakamura}\ \emph {et~al.}(1997)\citenamefont {Nakamura}, \citenamefont {Ikeda},\ and\ \citenamefont {Yamaguchi}}]{InSb_Nernst_1}%
  \BibitemOpen
  \bibfield  {author} {\bibinfo {author} {\bibfnamefont {H.}~\bibnamefont {Nakamura}}, \bibinfo {author} {\bibfnamefont {K.}~\bibnamefont {Ikeda}},\ and\ \bibinfo {author} {\bibfnamefont {S.}~\bibnamefont {Yamaguchi}},\ }\bibfield  {title} {\bibinfo {title} {{Transport Property and Energy Conversion of Nernst Elements in Strong Magnetic Field}},\ }\href {https://doi.org/10.2320/jinstmet1952.61.12_1318} {\bibfield  {journal} {\bibinfo  {journal} {Journal of the Japan Institute of Metals}\ }\textbf {\bibinfo {volume} {61}},\ \bibinfo {pages} {1318} (\bibinfo {year} {1997})}\BibitemShut {NoStop}%
\bibitem [{\citenamefont {Yang}\ \emph {et~al.}(2025)\citenamefont {Yang}, \citenamefont {Fang}, \citenamefont {Feng}, \citenamefont {Jiang}, \citenamefont {Zhang}, \citenamefont {Liu}, \citenamefont {Cheng}, \citenamefont {Yang}, \citenamefont {Li}, \citenamefont {Liang}, \citenamefont {Zheng}, \citenamefont {Deng}, \citenamefont {Qi},\ and\ \citenamefont {Liu}}]{Synt_Yang_Turbostratcic_Self_Heating}%
  \BibitemOpen
  \bibfield  {author} {\bibinfo {author} {\bibfnamefont {Y.}~\bibnamefont {Yang}}, \bibinfo {author} {\bibfnamefont {Y.}~\bibnamefont {Fang}}, \bibinfo {author} {\bibfnamefont {E.}~\bibnamefont {Feng}}, \bibinfo {author} {\bibfnamefont {W.}~\bibnamefont {Jiang}}, \bibinfo {author} {\bibfnamefont {X.}~\bibnamefont {Zhang}}, \bibinfo {author} {\bibfnamefont {L.}~\bibnamefont {Liu}}, \bibinfo {author} {\bibfnamefont {Y.}~\bibnamefont {Cheng}}, \bibinfo {author} {\bibfnamefont {F.}~\bibnamefont {Yang}}, \bibinfo {author} {\bibfnamefont {W.}~\bibnamefont {Li}}, \bibinfo {author} {\bibfnamefont {F.}~\bibnamefont {Liang}}, \bibinfo {author} {\bibfnamefont {K.}~\bibnamefont {Zheng}}, \bibinfo {author} {\bibfnamefont {B.}~\bibnamefont {Deng}}, \bibinfo {author} {\bibfnamefont {Y.}~\bibnamefont {Qi}},\ and\ \bibinfo {author} {\bibfnamefont {Z.}~\bibnamefont {Liu}},\ }\bibfield  {title} {\bibinfo {title} {Scalable, universal in situ self-heating chemical vapor deposition strategy for high-quality thick turbostratic
  graphene via combined twist–tilt configuration engineering},\ }\href {https://doi.org/10.1021/jacs.5c14727} {\bibfield  {journal} {\bibinfo  {journal} {Journal of the American Chemical Society}\ }\textbf {\bibinfo {volume} {147}},\ \bibinfo {pages} {43805} (\bibinfo {year} {2025})}\BibitemShut {NoStop}%
\bibitem [{\citenamefont {Garlow}\ \emph {et~al.}(2016)\citenamefont {Garlow}, \citenamefont {Barrett}, \citenamefont {Wu}, \citenamefont {Kisslinger}, \citenamefont {Zhu},\ and\ \citenamefont {Pulecio}}]{Synt_Garlow_TurbostraticMLG_Raman}%
  \BibitemOpen
  \bibfield  {author} {\bibinfo {author} {\bibfnamefont {J.~A.}\ \bibnamefont {Garlow}}, \bibinfo {author} {\bibfnamefont {L.~K.}\ \bibnamefont {Barrett}}, \bibinfo {author} {\bibfnamefont {L.}~\bibnamefont {Wu}}, \bibinfo {author} {\bibfnamefont {K.}~\bibnamefont {Kisslinger}}, \bibinfo {author} {\bibfnamefont {Y.}~\bibnamefont {Zhu}},\ and\ \bibinfo {author} {\bibfnamefont {J.~F.}\ \bibnamefont {Pulecio}},\ }\bibfield  {title} {\bibinfo {title} {{Large-Area Growth of Turbostratic Graphene on Ni(111) via Physical Vapor Deposition}},\ }\href {https://doi.org/10.1038/srep19804} {\bibfield  {journal} {\bibinfo  {journal} {Scientific Reports}\ }\textbf {\bibinfo {volume} {6}},\ \bibinfo {pages} {19804} (\bibinfo {year} {2016})}\BibitemShut {NoStop}%
\bibitem [{\citenamefont {Brzhezinskaya}\ \emph {et~al.}(2021)\citenamefont {Brzhezinskaya}, \citenamefont {Kononenko}, \citenamefont {Matveev}, \citenamefont {Zotov}, \citenamefont {Khodos}, \citenamefont {Levashov}, \citenamefont {Volkov}, \citenamefont {Bozhko}, \citenamefont {Chekmazov},\ and\ \citenamefont {Roshchupkin}}]{Brzhezinskaya2021}%
  \BibitemOpen
  \bibfield  {author} {\bibinfo {author} {\bibfnamefont {M.}~\bibnamefont {Brzhezinskaya}}, \bibinfo {author} {\bibfnamefont {O.}~\bibnamefont {Kononenko}}, \bibinfo {author} {\bibfnamefont {V.}~\bibnamefont {Matveev}}, \bibinfo {author} {\bibfnamefont {A.}~\bibnamefont {Zotov}}, \bibinfo {author} {\bibfnamefont {I.~I.}\ \bibnamefont {Khodos}}, \bibinfo {author} {\bibfnamefont {V.}~\bibnamefont {Levashov}}, \bibinfo {author} {\bibfnamefont {V.}~\bibnamefont {Volkov}}, \bibinfo {author} {\bibfnamefont {S.~I.}\ \bibnamefont {Bozhko}}, \bibinfo {author} {\bibfnamefont {S.~V.}\ \bibnamefont {Chekmazov}},\ and\ \bibinfo {author} {\bibfnamefont {D.}~\bibnamefont {Roshchupkin}},\ }\bibfield  {title} {\bibinfo {title} {{Engineering of Numerous Moir{\'{e}} Superlattices in Twisted Multilayer Graphene for Twistronics and Straintronics Applications}},\ }\href {https://doi.org/10.1021/acsnano.1c04286} {\bibfield  {journal} {\bibinfo  {journal} {ACS Nano}\ }\textbf {\bibinfo {volume} {15}},\ \bibinfo {pages} {12358}
  (\bibinfo {year} {2021})}\BibitemShut {NoStop}%
\bibitem [{\citenamefont {Kononenko}\ \emph {et~al.}(2022)\citenamefont {Kononenko}, \citenamefont {Brzhezinskaya}, \citenamefont {Zotov}, \citenamefont {Korepanov}, \citenamefont {Levashov}, \citenamefont {Matveev},\ and\ \citenamefont {Roshchupkin}}]{Kononenko2022}%
  \BibitemOpen
  \bibfield  {author} {\bibinfo {author} {\bibfnamefont {O.}~\bibnamefont {Kononenko}}, \bibinfo {author} {\bibfnamefont {M.}~\bibnamefont {Brzhezinskaya}}, \bibinfo {author} {\bibfnamefont {A.}~\bibnamefont {Zotov}}, \bibinfo {author} {\bibfnamefont {V.}~\bibnamefont {Korepanov}}, \bibinfo {author} {\bibfnamefont {V.}~\bibnamefont {Levashov}}, \bibinfo {author} {\bibfnamefont {V.}~\bibnamefont {Matveev}},\ and\ \bibinfo {author} {\bibfnamefont {D.}~\bibnamefont {Roshchupkin}},\ }\bibfield  {title} {\bibinfo {title} {{Influence of numerous Moir{\'{e}} superlattices on transport properties of twisted multilayer graphene}},\ }\href {https://doi.org/10.1016/j.carbon.2022.03.033} {\bibfield  {journal} {\bibinfo  {journal} {Carbon}\ }\textbf {\bibinfo {volume} {194}},\ \bibinfo {pages} {52} (\bibinfo {year} {2022})}\BibitemShut {NoStop}%
\bibitem [{\citenamefont {Li}\ \emph {et~al.}(2016)\citenamefont {Li}, \citenamefont {Zhang}, \citenamefont {He}, \citenamefont {Ren}, \citenamefont {Cheng},\ and\ \citenamefont {Zhang}}]{Li_GMR_graphitic_foam}%
  \BibitemOpen
  \bibfield  {author} {\bibinfo {author} {\bibfnamefont {P.}~\bibnamefont {Li}}, \bibinfo {author} {\bibfnamefont {Q.}~\bibnamefont {Zhang}}, \bibinfo {author} {\bibfnamefont {X.}~\bibnamefont {He}}, \bibinfo {author} {\bibfnamefont {W.}~\bibnamefont {Ren}}, \bibinfo {author} {\bibfnamefont {H.-M.}\ \bibnamefont {Cheng}},\ and\ \bibinfo {author} {\bibfnamefont {X.-x.}\ \bibnamefont {Zhang}},\ }\bibfield  {title} {\bibinfo {title} {Spatial mobility fluctuation induced giant linear magnetoresistance in multilayered graphene foam},\ }\href {https://doi.org/10.1103/PhysRevB.94.045402} {\bibfield  {journal} {\bibinfo  {journal} {Phys. Rev. B}\ }\textbf {\bibinfo {volume} {94}},\ \bibinfo {pages} {045402} (\bibinfo {year} {2016})}\BibitemShut {NoStop}%
\bibitem [{\citenamefont {Zhang}\ \emph {et~al.}(2024)\citenamefont {Zhang}, \citenamefont {Xie}, \citenamefont {Yang}, \citenamefont {Wu}, \citenamefont {Lu}, \citenamefont {Hu}, \citenamefont {Ding}, \citenamefont {He}, \citenamefont {Dong}, \citenamefont {Wang} \emph {et~al.}}]{MR_spirals}%
  \BibitemOpen
  \bibfield  {author} {\bibinfo {author} {\bibfnamefont {Y.}~\bibnamefont {Zhang}}, \bibinfo {author} {\bibfnamefont {B.}~\bibnamefont {Xie}}, \bibinfo {author} {\bibfnamefont {Y.}~\bibnamefont {Yang}}, \bibinfo {author} {\bibfnamefont {Y.}~\bibnamefont {Wu}}, \bibinfo {author} {\bibfnamefont {X.}~\bibnamefont {Lu}}, \bibinfo {author} {\bibfnamefont {Y.}~\bibnamefont {Hu}}, \bibinfo {author} {\bibfnamefont {Y.}~\bibnamefont {Ding}}, \bibinfo {author} {\bibfnamefont {J.}~\bibnamefont {He}}, \bibinfo {author} {\bibfnamefont {P.}~\bibnamefont {Dong}}, \bibinfo {author} {\bibfnamefont {J.}~\bibnamefont {Wang}}, \emph {et~al.},\ }\bibfield  {title} {\bibinfo {title} {Extremely large magnetoresistance in twisted intertwined graphene spirals},\ }\href {https://doi.org/10.1038/s41467-024-50456-0} {\bibfield  {journal} {\bibinfo  {journal} {Nature Communications}\ }\textbf {\bibinfo {volume} {15}},\ \bibinfo {pages} {6120} (\bibinfo {year} {2024})}\BibitemShut {NoStop}%
\bibitem [{\citenamefont {Wu}\ \emph {et~al.}(2016)\citenamefont {Wu}, \citenamefont {Wang}, \citenamefont {Li}, \citenamefont {Peng},\ and\ \citenamefont {Tan}}]{Synt_Wu_Raman_TMLG}%
  \BibitemOpen
  \bibfield  {author} {\bibinfo {author} {\bibfnamefont {J.-B.}\ \bibnamefont {Wu}}, \bibinfo {author} {\bibfnamefont {H.}~\bibnamefont {Wang}}, \bibinfo {author} {\bibfnamefont {X.-L.}\ \bibnamefont {Li}}, \bibinfo {author} {\bibfnamefont {H.}~\bibnamefont {Peng}},\ and\ \bibinfo {author} {\bibfnamefont {P.-H.}\ \bibnamefont {Tan}},\ }\bibfield  {title} {\bibinfo {title} {{Raman spectroscopic characterization of stacking configuration and interlayer coupling of twisted multilayer graphene grown by chemical vapor deposition}},\ }\href {https://doi.org/10.1016/j.carbon.2016.09.006} {\bibfield  {journal} {\bibinfo  {journal} {Carbon}\ }\textbf {\bibinfo {volume} {110}},\ \bibinfo {pages} {225} (\bibinfo {year} {2016})}\BibitemShut {NoStop}%
\bibitem [{\citenamefont {Parish}\ and\ \citenamefont {Littlewood}(2003)}]{EMT_magnetoresistance_2}%
  \BibitemOpen
  \bibfield  {author} {\bibinfo {author} {\bibfnamefont {M.~M.}\ \bibnamefont {Parish}}\ and\ \bibinfo {author} {\bibfnamefont {P.~B.}\ \bibnamefont {Littlewood}},\ }\bibfield  {title} {\bibinfo {title} {{Non-saturating magnetoresistance in heavily disordered semiconductors}},\ }\href {https://doi.org/10.1038/nature02073} {\bibfield  {journal} {\bibinfo  {journal} {Nature}\ }\textbf {\bibinfo {volume} {426}},\ \bibinfo {pages} {162} (\bibinfo {year} {2003})}\BibitemShut {NoStop}%
\bibitem [{\citenamefont {Guttal}\ and\ \citenamefont {Stroud}(2005)}]{EMT_magnetoresistance_1}%
  \BibitemOpen
  \bibfield  {author} {\bibinfo {author} {\bibfnamefont {V.}~\bibnamefont {Guttal}}\ and\ \bibinfo {author} {\bibfnamefont {D.}~\bibnamefont {Stroud}},\ }\bibfield  {title} {\bibinfo {title} {{Model for a macroscopically disordered conductor with an exactly linear high-field magnetoresistance}},\ }\href {https://doi.org/10.1103/PhysRevB.71.201304} {\bibfield  {journal} {\bibinfo  {journal} {Physical Review B - Condensed Matter and Materials Physics}\ }\textbf {\bibinfo {volume} {71}},\ \bibinfo {pages} {1} (\bibinfo {year} {2005})}\BibitemShut {NoStop}%
\bibitem [{\citenamefont {Guttal}\ and\ \citenamefont {Stroud}(2006)}]{EMT_hall_effect}%
  \BibitemOpen
  \bibfield  {author} {\bibinfo {author} {\bibfnamefont {V.}~\bibnamefont {Guttal}}\ and\ \bibinfo {author} {\bibfnamefont {D.}~\bibnamefont {Stroud}},\ }\bibfield  {title} {\bibinfo {title} {{Nonsaturating magnetoresistance and Hall coefficient reversal in a model composite semiconductor}},\ }\href {https://doi.org/10.1103/PhysRevB.73.085202} {\bibfield  {journal} {\bibinfo  {journal} {Physical Review B - Condensed Matter and Materials Physics}\ }\textbf {\bibinfo {volume} {73}},\ \bibinfo {pages} {085202} (\bibinfo {year} {2006})}\BibitemShut {NoStop}%
\bibitem [{\citenamefont {Batlle~Porro}\ \emph {et~al.}(2025)\citenamefont {Batlle~Porro}, \citenamefont {C{\u{a}}lug{\u{a}}ru}, \citenamefont {Hu}, \citenamefont {Krishna~Kumar}, \citenamefont {Hesp}, \citenamefont {Watanabe}, \citenamefont {Taniguchi}, \citenamefont {Bernevig}, \citenamefont {Stepanov},\ and\ \citenamefont {Koppens}}]{Stepanov_TTG_Thermopower}%
  \BibitemOpen
  \bibfield  {author} {\bibinfo {author} {\bibfnamefont {S.}~\bibnamefont {Batlle~Porro}}, \bibinfo {author} {\bibfnamefont {D.}~\bibnamefont {C{\u{a}}lug{\u{a}}ru}}, \bibinfo {author} {\bibfnamefont {H.}~\bibnamefont {Hu}}, \bibinfo {author} {\bibfnamefont {R.}~\bibnamefont {Krishna~Kumar}}, \bibinfo {author} {\bibfnamefont {N.~C.}\ \bibnamefont {Hesp}}, \bibinfo {author} {\bibfnamefont {K.}~\bibnamefont {Watanabe}}, \bibinfo {author} {\bibfnamefont {T.}~\bibnamefont {Taniguchi}}, \bibinfo {author} {\bibfnamefont {B.~A.}\ \bibnamefont {Bernevig}}, \bibinfo {author} {\bibfnamefont {P.}~\bibnamefont {Stepanov}},\ and\ \bibinfo {author} {\bibfnamefont {F.~H.}\ \bibnamefont {Koppens}},\ }\bibfield  {title} {\bibinfo {title} {Photovoltage microscopy of symmetrically twisted trilayer graphene},\ }\href {https://doi.org/10.1038/s41567-025-03071-9} {\bibfield  {journal} {\bibinfo  {journal} {Nature Physics}\ }\textbf {\bibinfo {volume} {21}},\ \bibinfo {pages} {1934} (\bibinfo {year} {2025})}\BibitemShut {NoStop}%
\bibitem [{\citenamefont {Sunku}\ \emph {et~al.}(2021)\citenamefont {Sunku}, \citenamefont {Halbertal}, \citenamefont {Stauber}, \citenamefont {Chen}, \citenamefont {McLeod}, \citenamefont {Rikhter}, \citenamefont {Berkowitz}, \citenamefont {Lo}, \citenamefont {Gonzalez-Acevedo}, \citenamefont {Hone}, \citenamefont {Dean}, \citenamefont {Fogler},\ and\ \citenamefont {Basov}}]{Basov_LocalThermopower}%
  \BibitemOpen
  \bibfield  {author} {\bibinfo {author} {\bibfnamefont {S.~S.}\ \bibnamefont {Sunku}}, \bibinfo {author} {\bibfnamefont {D.}~\bibnamefont {Halbertal}}, \bibinfo {author} {\bibfnamefont {T.}~\bibnamefont {Stauber}}, \bibinfo {author} {\bibfnamefont {S.}~\bibnamefont {Chen}}, \bibinfo {author} {\bibfnamefont {A.~S.}\ \bibnamefont {McLeod}}, \bibinfo {author} {\bibfnamefont {A.}~\bibnamefont {Rikhter}}, \bibinfo {author} {\bibfnamefont {M.~E.}\ \bibnamefont {Berkowitz}}, \bibinfo {author} {\bibfnamefont {C.~F.~B.}\ \bibnamefont {Lo}}, \bibinfo {author} {\bibfnamefont {D.~E.}\ \bibnamefont {Gonzalez-Acevedo}}, \bibinfo {author} {\bibfnamefont {J.~C.}\ \bibnamefont {Hone}}, \bibinfo {author} {\bibfnamefont {C.~R.}\ \bibnamefont {Dean}}, \bibinfo {author} {\bibfnamefont {M.~M.}\ \bibnamefont {Fogler}},\ and\ \bibinfo {author} {\bibfnamefont {D.~N.}\ \bibnamefont {Basov}},\ }\bibfield  {title} {\bibinfo {title} {{Hyperbolic enhancement of photocurrent patterns in minimally twisted bilayer graphene}},\ }\href
  {https://doi.org/10.1038/s41467-021-21792-2} {\bibfield  {journal} {\bibinfo  {journal} {Nature Communications}\ }\textbf {\bibinfo {volume} {12}},\ \bibinfo {pages} {1641} (\bibinfo {year} {2021})}\BibitemShut {NoStop}%
\bibitem [{Note1()}]{Note1}%
  \BibitemOpen
  \bibinfo {note} {The size of the focused beam was estimated in the independent series of experiments, where the photovoltage maps of miniature photodetectors were recorded and fitted with Gaussian function $V_{\protect \rm ph}(\protect \bf r) \propto \exp \{-|{\protect \bf r}|^2/2\sigma ^2\}$. Best fits yield $\sigma \approx 20$ $\mu $m, thus the spatial distribution of the lase power density can be presented as $p({\protect \bf r}) = P_0 (2\pi \sigma ^2)^{-1} \exp \{-|{\protect \bf r}|^2/2\sigma ^2\}$}\BibitemShut {NoStop}%
\bibitem [{\citenamefont {Xia}\ \emph {et~al.}(2009)\citenamefont {Xia}, \citenamefont {Mueller}, \citenamefont {Golizadeh-Mojarad}, \citenamefont {Freitage}, \citenamefont {Lin}, \citenamefont {Tsang}, \citenamefont {Perebeinos},\ and\ \citenamefont {Avouris}}]{Xia_photocurrent_imaging}%
  \BibitemOpen
  \bibfield  {author} {\bibinfo {author} {\bibfnamefont {F.}~\bibnamefont {Xia}}, \bibinfo {author} {\bibfnamefont {T.}~\bibnamefont {Mueller}}, \bibinfo {author} {\bibfnamefont {R.}~\bibnamefont {Golizadeh-Mojarad}}, \bibinfo {author} {\bibfnamefont {M.}~\bibnamefont {Freitage}}, \bibinfo {author} {\bibfnamefont {Y.~M.}\ \bibnamefont {Lin}}, \bibinfo {author} {\bibfnamefont {J.}~\bibnamefont {Tsang}}, \bibinfo {author} {\bibfnamefont {V.}~\bibnamefont {Perebeinos}},\ and\ \bibinfo {author} {\bibfnamefont {P.}~\bibnamefont {Avouris}},\ }\bibfield  {title} {\bibinfo {title} {{Photocurrent imaging and efficient photon detection in a graphene transistor}},\ }\href {https://doi.org/10.1021/nl8033812} {\bibfield  {journal} {\bibinfo  {journal} {Nano Letters}\ }\textbf {\bibinfo {volume} {9}},\ \bibinfo {pages} {1039} (\bibinfo {year} {2009})}\BibitemShut {NoStop}%
\bibitem [{\citenamefont {Xu}\ \emph {et~al.}(2010)\citenamefont {Xu}, \citenamefont {Gabor}, \citenamefont {Alden}, \citenamefont {{Van Der Zande}},\ and\ \citenamefont {McEuen}}]{Gabor_junction_PTE}%
  \BibitemOpen
  \bibfield  {author} {\bibinfo {author} {\bibfnamefont {X.}~\bibnamefont {Xu}}, \bibinfo {author} {\bibfnamefont {N.~M.}\ \bibnamefont {Gabor}}, \bibinfo {author} {\bibfnamefont {J.~S.}\ \bibnamefont {Alden}}, \bibinfo {author} {\bibfnamefont {A.~M.}\ \bibnamefont {{Van Der Zande}}},\ and\ \bibinfo {author} {\bibfnamefont {P.~L.}\ \bibnamefont {McEuen}},\ }\bibfield  {title} {\bibinfo {title} {{Photo-thermoelectric effect at a graphene interface junction}},\ }\href {https://doi.org/10.1021/nl903451y} {\bibfield  {journal} {\bibinfo  {journal} {Nano Letters}\ }\textbf {\bibinfo {volume} {10}},\ \bibinfo {pages} {562} (\bibinfo {year} {2010})},\ \Eprint {https://arxiv.org/abs/0907.3173} {0907.3173} \BibitemShut {NoStop}%
\bibitem [{\citenamefont {Woessner}\ \emph {et~al.}(2016)\citenamefont {Woessner}, \citenamefont {Alonso-Gonz{\'{a}}lez}, \citenamefont {Lundeberg}, \citenamefont {Gao}, \citenamefont {Barrios-Vargas}, \citenamefont {Navickaite}, \citenamefont {Ma}, \citenamefont {Janner}, \citenamefont {Watanabe}, \citenamefont {Cummings}, \citenamefont {Taniguchi}, \citenamefont {Pruneri}, \citenamefont {Roche}, \citenamefont {Jarillo-Herrero}, \citenamefont {Hone}, \citenamefont {Hillenbrand},\ and\ \citenamefont {Koppens}}]{Woessner_nanoscopy}%
  \BibitemOpen
  \bibfield  {author} {\bibinfo {author} {\bibfnamefont {A.}~\bibnamefont {Woessner}}, \bibinfo {author} {\bibfnamefont {P.}~\bibnamefont {Alonso-Gonz{\'{a}}lez}}, \bibinfo {author} {\bibfnamefont {M.~B.}\ \bibnamefont {Lundeberg}}, \bibinfo {author} {\bibfnamefont {Y.}~\bibnamefont {Gao}}, \bibinfo {author} {\bibfnamefont {J.~E.}\ \bibnamefont {Barrios-Vargas}}, \bibinfo {author} {\bibfnamefont {G.}~\bibnamefont {Navickaite}}, \bibinfo {author} {\bibfnamefont {Q.}~\bibnamefont {Ma}}, \bibinfo {author} {\bibfnamefont {D.}~\bibnamefont {Janner}}, \bibinfo {author} {\bibfnamefont {K.}~\bibnamefont {Watanabe}}, \bibinfo {author} {\bibfnamefont {A.~W.}\ \bibnamefont {Cummings}}, \bibinfo {author} {\bibfnamefont {T.}~\bibnamefont {Taniguchi}}, \bibinfo {author} {\bibfnamefont {V.}~\bibnamefont {Pruneri}}, \bibinfo {author} {\bibfnamefont {S.}~\bibnamefont {Roche}}, \bibinfo {author} {\bibfnamefont {P.}~\bibnamefont {Jarillo-Herrero}}, \bibinfo {author} {\bibfnamefont {J.}~\bibnamefont {Hone}}, \bibinfo {author}
  {\bibfnamefont {R.}~\bibnamefont {Hillenbrand}},\ and\ \bibinfo {author} {\bibfnamefont {F.~H.~L.}\ \bibnamefont {Koppens}},\ }\bibfield  {title} {\bibinfo {title} {{Near-field photocurrent nanoscopy on bare and encapsulated graphene}},\ }\href {https://doi.org/10.1038/ncomms10783} {\bibfield  {journal} {\bibinfo  {journal} {Nature Communications}\ }\textbf {\bibinfo {volume} {7}},\ \bibinfo {pages} {10783} (\bibinfo {year} {2016})},\ \Eprint {https://arxiv.org/abs/1508.07864} {1508.07864} \BibitemShut {NoStop}%
\bibitem [{\citenamefont {Song}\ and\ \citenamefont {Levitov}(2014)}]{Levitov_Osng_SHockley_Ramo}%
  \BibitemOpen
  \bibfield  {author} {\bibinfo {author} {\bibfnamefont {J.~C.~W.}\ \bibnamefont {Song}}\ and\ \bibinfo {author} {\bibfnamefont {L.~S.}\ \bibnamefont {Levitov}},\ }\bibfield  {title} {\bibinfo {title} {{Shockley-Ramo theorem and long-range photocurrent response in gapless materials}},\ }\href {https://doi.org/10.1103/PhysRevB.90.075415} {\bibfield  {journal} {\bibinfo  {journal} {Physical Review B}\ }\textbf {\bibinfo {volume} {90}},\ \bibinfo {pages} {075415} (\bibinfo {year} {2014})},\ \Eprint {https://arxiv.org/abs/1112.5654} {1112.5654} \BibitemShut {NoStop}%
\bibitem [{\citenamefont {Rezaei}\ and\ \citenamefont {Schindler}(2024)}]{ABA_graphene_nernst}%
  \BibitemOpen
  \bibfield  {author} {\bibinfo {author} {\bibfnamefont {S.~E.}\ \bibnamefont {Rezaei}}\ and\ \bibinfo {author} {\bibfnamefont {P.}~\bibnamefont {Schindler}},\ }\bibfield  {title} {\bibinfo {title} {{Revealing large room-temperature Nernst coefficients in 2D materials by first-principles modeling}},\ }\href {https://doi.org/10.1039/d3nr06127b} {\bibfield  {journal} {\bibinfo  {journal} {Nanoscale}\ }\textbf {\bibinfo {volume} {16}},\ \bibinfo {pages} {6142} (\bibinfo {year} {2024})}\BibitemShut {NoStop}%
\bibitem [{\citenamefont {Jernigan}\ \emph {et~al.}(2009)\citenamefont {Jernigan}, \citenamefont {VanMil}, \citenamefont {Tedesco}, \citenamefont {Tischler}, \citenamefont {Glaser}, \citenamefont {Davidson}, \citenamefont {Campbell},\ and\ \citenamefont {Gaskill}}]{Gaskill_NL_Infrared_C_face}%
  \BibitemOpen
  \bibfield  {author} {\bibinfo {author} {\bibfnamefont {G.~G.}\ \bibnamefont {Jernigan}}, \bibinfo {author} {\bibfnamefont {B.~L.}\ \bibnamefont {VanMil}}, \bibinfo {author} {\bibfnamefont {J.~L.}\ \bibnamefont {Tedesco}}, \bibinfo {author} {\bibfnamefont {J.~G.}\ \bibnamefont {Tischler}}, \bibinfo {author} {\bibfnamefont {E.~R.}\ \bibnamefont {Glaser}}, \bibinfo {author} {\bibfnamefont {I.}~\bibnamefont {Davidson}, \bibfnamefont {Anthony}}, \bibinfo {author} {\bibfnamefont {P.~M.}\ \bibnamefont {Campbell}},\ and\ \bibinfo {author} {\bibfnamefont {D.~K.}\ \bibnamefont {Gaskill}},\ }\bibfield  {title} {\bibinfo {title} {Comparison of epitaxial graphene on si-face and c-face 4h sic formed by ultrahigh vacuum and rf furnace production},\ }\href {https://doi.org/10.1021/nl900803z} {\bibfield  {journal} {\bibinfo  {journal} {Nano Letters}\ }\textbf {\bibinfo {volume} {9}},\ \bibinfo {pages} {2605} (\bibinfo {year} {2009})}\BibitemShut {NoStop}%
\bibitem [{\citenamefont {Alymov}\ \emph {et~al.}(2018)\citenamefont {Alymov}, \citenamefont {Vyurkov}, \citenamefont {Ryzhii}, \citenamefont {Satou},\ and\ \citenamefont {Svintsov}}]{Alymov2018a}%
  \BibitemOpen
  \bibfield  {author} {\bibinfo {author} {\bibfnamefont {G.}~\bibnamefont {Alymov}}, \bibinfo {author} {\bibfnamefont {V.}~\bibnamefont {Vyurkov}}, \bibinfo {author} {\bibfnamefont {V.}~\bibnamefont {Ryzhii}}, \bibinfo {author} {\bibfnamefont {A.}~\bibnamefont {Satou}},\ and\ \bibinfo {author} {\bibfnamefont {D.}~\bibnamefont {Svintsov}},\ }\bibfield  {title} {\bibinfo {title} {{Auger recombination in Dirac materials: A tangle of many-body effects}},\ }\href {https://doi.org/10.1103/PhysRevB.97.205411} {\bibfield  {journal} {\bibinfo  {journal} {Physical Review B}\ }\textbf {\bibinfo {volume} {97}},\ \bibinfo {pages} {205411} (\bibinfo {year} {2018})},\ \Eprint {https://arxiv.org/abs/1709.09015} {1709.09015} \BibitemShut {NoStop}%
\bibitem [{\citenamefont {Du}\ \emph {et~al.}(2023)\citenamefont {Du}, \citenamefont {Xie}, \citenamefont {Yin}, \citenamefont {Sun}, \citenamefont {Wang}, \citenamefont {Liu}, \citenamefont {Qi}, \citenamefont {Cai}, \citenamefont {Bi}, \citenamefont {Xiao}, \citenamefont {Chen}, \citenamefont {Shen}, \citenamefont {Yin},\ and\ \citenamefont {Zheng}}]{AR_multilayers}%
  \BibitemOpen
  \bibfield  {author} {\bibinfo {author} {\bibfnamefont {S.}~\bibnamefont {Du}}, \bibinfo {author} {\bibfnamefont {H.}~\bibnamefont {Xie}}, \bibinfo {author} {\bibfnamefont {J.}~\bibnamefont {Yin}}, \bibinfo {author} {\bibfnamefont {Y.}~\bibnamefont {Sun}}, \bibinfo {author} {\bibfnamefont {Q.}~\bibnamefont {Wang}}, \bibinfo {author} {\bibfnamefont {H.}~\bibnamefont {Liu}}, \bibinfo {author} {\bibfnamefont {W.}~\bibnamefont {Qi}}, \bibinfo {author} {\bibfnamefont {C.}~\bibnamefont {Cai}}, \bibinfo {author} {\bibfnamefont {G.}~\bibnamefont {Bi}}, \bibinfo {author} {\bibfnamefont {D.}~\bibnamefont {Xiao}}, \bibinfo {author} {\bibfnamefont {W.}~\bibnamefont {Chen}}, \bibinfo {author} {\bibfnamefont {X.}~\bibnamefont {Shen}}, \bibinfo {author} {\bibfnamefont {W.-Y.}\ \bibnamefont {Yin}},\ and\ \bibinfo {author} {\bibfnamefont {R.}~\bibnamefont {Zheng}},\ }\bibfield  {title} {\bibinfo {title} {Giant hot electron thermalization via stacking of graphene layers},\ }\href
  {https://doi.org/https://doi.org/10.1016/j.carbon.2022.12.017} {\bibfield  {journal} {\bibinfo  {journal} {Carbon}\ }\textbf {\bibinfo {volume} {203}},\ \bibinfo {pages} {835} (\bibinfo {year} {2023})}\BibitemShut {NoStop}%
\bibitem [{\citenamefont {Hass}\ \emph {et~al.}(2008)\citenamefont {Hass}, \citenamefont {de~Heer},\ and\ \citenamefont {Conrad}}]{MLG_on_SiC}%
  \BibitemOpen
  \bibfield  {author} {\bibinfo {author} {\bibfnamefont {J.}~\bibnamefont {Hass}}, \bibinfo {author} {\bibfnamefont {W.~A.}\ \bibnamefont {de~Heer}},\ and\ \bibinfo {author} {\bibfnamefont {E.~H.}\ \bibnamefont {Conrad}},\ }\bibfield  {title} {\bibinfo {title} {The growth and morphology of epitaxial multilayer graphene},\ }\href {https://doi.org/10.1088/0953-8984/20/32/323202} {\bibfield  {journal} {\bibinfo  {journal} {Journal of Physics: Condensed Matter}\ }\textbf {\bibinfo {volume} {20}},\ \bibinfo {pages} {323202} (\bibinfo {year} {2008})}\BibitemShut {NoStop}%
\bibitem [{\citenamefont {Balagurov}(1986)}]{balagurov1986theory}%
  \BibitemOpen
  \bibfield  {author} {\bibinfo {author} {\bibfnamefont {B.~Y.}\ \bibnamefont {Balagurov}},\ }\bibfield  {title} {\bibinfo {title} {To the theory of thermoelectric properties of bicomponent media},\ }\href {https://www.mathnet.ru/php/archive.phtml?wshow=paper&jrnid=phts&paperid=298&option_lang=eng} {\bibfield  {journal} {\bibinfo  {journal} {Fizika i Tekhnika Poluprovodnikov}\ }\textbf {\bibinfo {volume} {20}},\ \bibinfo {pages} {1276} (\bibinfo {year} {1986})}\BibitemShut {NoStop}%
\bibitem [{\citenamefont {Fishchuk}(1995)}]{Fishchuk_Nernst_Disordered}%
  \BibitemOpen
  \bibfield  {author} {\bibinfo {author} {\bibfnamefont {I.~I.}\ \bibnamefont {Fishchuk}},\ }\bibfield  {title} {\bibinfo {title} {{Theory of the Thermopower and Nernst Effect in Random Two‐Component Solid Systems}},\ }\href {https://doi.org/10.1002/pssb.2221900225} {\bibfield  {journal} {\bibinfo  {journal} {physica status solidi (b)}\ }\textbf {\bibinfo {volume} {190}},\ \bibinfo {pages} {545} (\bibinfo {year} {1995})}\BibitemShut {NoStop}%
\end{thebibliography}%

\end{document}